\documentclass{article}

\usepackage{cite,amssymb,amsmath,epsf,graphics
}

\def\1ad{\mbox{\normalsize $^1$}}
\def\2ad{\mbox{\normalsize $^2$}}
\def\3ad{\mbox{\normalsize $^3$}}
\def\4ad{\mbox{\normalsize $^4$}}
\def\5ad{\mbox{\normalsize $^5$}}
\def\6ad{\mbox{\normalsize $^6$}}
\def\7ad{\mbox{\normalsize $^7$}}
\def\8ad{\mbox{\normalsize $^8$}}
\def\makefront{
\vspace*{1cm}\begin{center}
\def\sp{
\renewcommand{\thefootnote}{\fnsymbol{footnote}}
\footnote[4]{corresponding author : \email_speaker}
\renewcommand{\thefootnote}{\arabic{footnote}}
}
\def\newtitleline{\\ \vskip 5pt}
{\Large\bf\titleline}\\
\vskip 1truecm
{\large\bf\authors}\\
\vskip 5truemm
\addresses
\end{center}
\vskip 1truecm
{\bf Abstract:}
\abstracttext
\vskip 1truecm
}
\numberwithin{equation}{section}
\newcommand{\al}{\alpha}

\newcommand{\ga}{\gamma}

\newcommand{\eps}{\epsilon}
\newcommand{\om}{\omega}

\newcommand{\vth}{\vartheta}

\newcommand{\Om}{\Omega}

\newcommand{\cN}{{\cal N}}

\newcommand{\beq}{\begin{equation}}
\newcommand{\eeq}{\end{equation}}
\newcommand{\bea}{\begin{eqnarray}}
\newcommand{\eea}{\end{eqnarray}}
\newcommand{\bean}{\begin{eqnarray*}}
\newcommand{\eean}{\end{eqnarray*}}
\newcommand{\non}{\nonumber}
\newcommand{\Bz}{{\mathbb Z}}
\newcommand{\Br}{{\mathbb R}}

\newcommand{\ident}{1 \hspace{-1mm} 1}

\newcommand{\tga}{\tilde{\gamma}}

\newcommand{\he}{\hat{e}}

\newcommand{\hD}{\hat{D}}

\newcommand{\hG}{\hat{G}}

\newcommand{\hga}{\hat{\gamma}}

\newcommand{\haom}{\hat{\omega}}

\newcommand{\hPsi}{\hat{\Psi}}

\newcommand{\im}{{\rm Im}}
\newcommand{\re}{{\rm Re}}

\begin{document}
\setcounter{page}{0}
\rightline{hep-th/0303127}
\rightline{CPHT-RR 011.0303}
\rightline{IHES/P/03/17}
\vskip 2cm

\def\titleline{
Supersymmetric M-theory compactifications  %
with fluxes \newtitleline
on seven-manifolds and $G$-structures                        
}
\def\email_speaker{
{\tt 
%
%
}}
\def\authors{
%
%
%
%
%
  Peter Kaste\1ad, Ruben Minasian\1ad and Alessandro Tomasiello\2ad
}
\def\addresses{
%
%
%
%
\1ad
Centre de Physique Th{\'e}orique, Ecole 
Polytechnique\\
Unit{\'e} mixte du CNRS et de l'EP, UMR 7644\\ 
91128 Palaiseau Cedex, France\\
\vskip.3cm
\2ad
I.H.E.S.,
Le Bois-Marie, Bures-sur-Yvette, 91440 France
}
\def\abstracttext{
%
%
We consider Minkowski compactifications of M-theory on generic
seven-dimensional manifolds. After analyzing the conditions on 
the four-form flux,
we establish a set of relations between the 
components of the intrinsic torsion of the internal manifold and the
components of the four-form flux needed for preserving supersymmetry.
The existence of two nowhere vanishing vectors on any seven-manifold with
G$_2$ structure plays a crucial role in our analysis, leading to
the possibility of four-dimensional compactifications with $\cN=1$ and
$\cN=2$ supersymmetry.  

}
\large
\makefront

\vfill
\begin{flushleft}
{\today}\\
\end{flushleft}
\thispagestyle{empty}

\section{Introduction}

String theory compactifications with background fluxes 
are an old subject of study, but in spite of their physical interest
they are not yet as well understood as their purely geometric counterparts.
Different directions of research on the more general task of understanding 
supersymmetric solutions have recently been focusing on the language and 
techniques of $G$-structures. So far this has been applied to Neveu-Schwarz 
three-form \cite{gmpw, gmw, glmw, mgH} or to the two-form flux in type IIA 
\cite{kmpt,kmptt}. It is obviously interesting to extend these methods to 
Ramond-Ramond fluxes. In type IIA these different fluxes  
get organized in terms of the four-form flux of M-theory.
In this paper we therefore reconsider compactifications of M-theory to four
dimensions in the presence of background four-form fluxes and analyze
the conditions under which the vacuum preserves $\cN=1$
supersymmetry in four dimensions.\footnote{G-structures have also been 
applied \cite{gp} 
to M-theory to classify all possible supersymmetric solutions 
in 11 dimensions, without reference to compactifications. Our use of SU(3)
structures in seven 
dimensions somewhat parallels the one in \cite{gp} 
of SU(5) structures in eleven.}

The traditional approach has been so far 
largely based on Ans{\"a}tze for the conserved spinor, the metric and the 
background fluxes \cite{cr, dnw, as, bb7, ali, bj}.  An incomplete list of closely 
related M-brane solutions is \cite{fs, cco, mt}.
The idea is that $G$-structures provide instead an organizing
principle and help to draw more general conclusions; much as in the purely
geometrical case, where possible internal metrics for compactifications have
long since been classified using the concept of holonomy. In that case the lack
of explicit expressions for the metric has sometimes been largely compensated
by the amount of mathematical results known about them, and one can hope that
this happy story repeats here to some extent. We hasten to add that 
compact nonsingular seven-manifolds are subject here 
to the usual simple no-go arguments coming from leading terms in the 
equations of motion \cite{mn,dhs,bb7}, which remain untouched by our analysis;
we are adding nothing here to the usual strategies to avoid this argument,
such as invoking higher-derivatives terms (not fully 
under control however as of this writing) and/or sources.
 
We look instead directly at supersymmetry. We can
take in general the eleven-dimensional spinors preserved by supersymmetry as
\begin{equation}
  \label{eq:11spinor}
\eps=\psi_+ \otimes \vth_+ + \psi_- \otimes \vth_- \ ,
\end{equation}
where $\psi_\pm=\psi_\mp^*$ are chiral spinors of 
opposite chiralities in four dimensions, and $\vth_+= \vth^*_-$ 
are some fixed seven-dimensional spinors due to 
eleven dimensional Majorana condition. Real and imaginary part of
$\vth_+$ define always an SU(3) structure on the seven-dimensional manifold
\footnote{In the degenerate case $\vth_+=\vth_-$ there is only one
spinor, which defines a G$_2$ structure. This case
does not however lead too far.}. The latter can also 
be reexpressed in a maybe more familiar terms using tensors $J, \Omega$ and a 
vector $v$ constructed as bilinears of the spinor, 
$\vth_\pm^\dagger \gamma_{i_1\ldots i_n} \vth_+$. We also find useful to 
think of it as a G$_2$ structure (defined by a real spinor 
$\vth \equiv {\rm Re}\{ e^{i \xi} \vth_+\}$). In this language, for example,
one recovers the Ansatz
\cite{dnw}
\begin{equation}
  \label{eq:dnw}
  \vth_{+} = |\vth_+| \left( 1+ v^a\ga_a \right)\vth
\end{equation}
as an inverse to the map $(\vth_\pm) \mapsto (v,\vth)$ just discussed. 
Indeed one might prefer to work with such an expression for the spinors, and 
make use of the way gamma matrices act on $\vth$, see eq. 
(\ref{g2.gamma.spin}). Our generalization of
the old spinor Ansatz (\ref{eq:dnw}) boils down to adding a phase.
This phase however carries an important geometrical information,
corresponding to the U(1) of SU(3) structures inside the G$_2$ structure.

As it happens, the presence of these two spinors on the manifold is no
loss of generality once the manifold is spin. A theorem 
\cite{fkms} which states, maybe somewhat surprisingly, 
that on a seven manifold a  spin structure implies an SU(2) structure (and 
thus in particular also a G$_2$ and an SU(3) structure) is the only
``input". On the other side, supposing that these
spinors, besides existing, are also supersymmetric, 
of course does lead to restrictions. These are of two types. There are 
constraints on the four-form $G$ and the warp factor $\Delta$, in which 
derivatives are only present in the form $d\Delta$ (and indirectly in the 
definition of $G= dC$) as will be seen in
(\ref{eq:dila}, \ref{eq:dila2}) and (\ref{eq:mono}).
As a comparison, let us recall that in M-theory 
compactifications on  four-folds \cite{bb8} 
there were primitivity constraints on $G$, 
whereas for two--form flux on manifolds of 
SU(3) structure a ``holomorphic monopole equation'' arose.
Then there are differential equations 
involving the tensors mentioned above:
\begin{equation}
  \label{eq:de}
  d(e^{2\Delta} v)=0\ , \qquad  d( e^{4 \Delta} J)= -2 \,e^{4 \Delta}* G\ ,
\qquad d(e^{3 \Delta} \Omega)=0\ .
\end{equation}

Remarkably --- or rather naturally, depending on the point of view --- these 
equations are indeed very similar to those found for NS three-form \cite{gmw}
or RR two-form \cite{kmpt}. This is no coincidence: 
the structure of these  equations is consistent with a brane interpretation
\cite{gp}. In particular, $J$ is said to be a {\sl generalized calibration}
for a five-brane that wraps a two-cycle inside $M$. 
This cycle can then be shown to minimize the energy of the brane, which takes
into account both the volume and the integral of the flux. In particular there
can be a non-trivial minimal energy cycle even in a trivial homology class.
This is somewhat in parallel with the fact for example 
that having SU(2) structure does not imply to have a non-trivial four-cycle,
as patently recalled by the above mentioned theorem about $G$-structures on 
seven-manifolds \cite{fkms}.

In terms of $G$-structures, equations (\ref{eq:de}) can be interpreted instead
as computing intrinsic torsions (as well as determining $G$ 
from the second one), quantities which measure the extent to 
which the manifold fails to have $G$-holonomy --- if one prefers, the amount 
of back-reaction. These objects are used to classify manifolds with 
$G$-structures. For instance, a weakly G$_2$ manifold is in this language 
simply
a manifold with G$_2$-structure whose intrinsic torsion is in the singlet
representation. Conformally G$_2$-holonomy 
manifolds have torsion in the vector representation, and so on. In our case,
what (\ref{eq:de}) teach us are SU(3) torsions, which although containing
more information, seem less useful for classification purposes. For this
reason, we  computed the G$_2$ torsions relative to the G$_2$ structure
defined by $\vth$ above. Although these do not contain all the information
about supersymmetry, they give simple necessary criteria for which 
G$_2$-structure manifolds can be used in presence of which fluxes. Finally,
a point on which our geometrical program fails is that 
the Bianchi identity (which
does not in general follow from supersymmetry) needs to be imposed separately,
as indeed it was done in all explicit examples based on Ans{\"a}tze. 
In general, the
intrinsic torsion can also be shown to satisfy differential equations and it
is a priori not
inconceivable that one might find cases in which this helps to solve Bianchi,
but this will not be settled here.

\section{Basics about $G$-structures}
\label{prelim}

Before starting with our analysis, we recall in this section some
basic concepts about $G$-structures. For details we refer the reader for 
example to \cite{j}.  

Consider an $n$-dimensional manifold $Y$, its tangent bundle $TY$ and its 
frame bundle $FY$. In general the latter has GL(n,$\Br$) as structure group,
namely this is the group in which its transition functions take value. It can
happen though that we can work with a smaller object: There can be a 
subbundle of $FY$, still principal (the structure group acts on
the fibers in the adjoint), but whose fibers are 
isomorphic to a smaller subgroup $G\subset$ GL(n,$\Br$). This subbundle
is called a $G$-structure on the manifold. The existence of such a
structure is a topological constraint. It implies that 
the structure group of $TY$ is reduced to $G$.

Tensors on $Y$ transform in some representation of the structure group 
GL(n,$\Br$). If a $G$-structure reduces this group to $G\subset$ GL(n,$\Br$),
then singlets may occur in the decomposition of the 
GL(n,$\Br$)-representation into irreducible $G$-representations, and these
singlets can be used to define alternatively the $G$-structure. 


A prime example is given by a Riemannian manifold. Existence
of a Riemannian metric $g$ on $Y$ allows one to define an O(n) subbundle 
of $FY$ defined by frames which are orthonormal, namely in which the metric
is written as $\delta_{ab}$. Orientability reduces to SO(n) and a spin 
structure is a reduction to Spin(n). Now, if
$G\subset$ Spin(n), 
also the spin-representation of SO(n) will contain
some singlets under $G$ corresponding to nowhere vanishing $G$-invariant
spinors on $Y$, that one can choose to have unit norm. So in these cases 
there is yet another way of characterizing the $G$-structure, through a 
$G$-invariant spinor. Obviously this is the case of interest for supersymmetry.
The $G$-invariant tensors mentioned above can be recovered from this spinor
$\vth$ as bilinears $\vth_\pm^\dagger \gamma_{i_1\ldots i_n} \vth_+$.
We will make all these concepts more explicit shortly for the case of SU(3)
and G$_2$ structures in seven dimensions. 

Obviously if there is a smaller $G$-structure it implies trivially the
existence  of a bigger one. Translating this in the tensor language, it is
interesting
to notice that this allows to recover the tensor characterizing the bigger
structure from the tensor characterizing the smaller structure. For example,
since G$_2 \subset $ SO(7) the existence of a G$_2$ invariant three-form allows
to find a metric associated to it --- this is the well-known formula 
for $g$ in terms  of $\Phi$. To go from a bigger to a smaller $G$ structure
is instead not obvious. If we now restrict our attention to dimension
seven we however find a surprise. In \cite{t} 
it had been shown that any compact, orientable
seven-manifold admits two linearly independent never vanishing
vector fields. This 
makes use of an index invariant for fields of 2-vectors on $n$-manifolds
analogous to the 
one used for simple vector fields. But, instead of being defined in 
$\pi_{n-1}(S^{n-1})=\Bz$, it is defined in  $\pi_{n-1}(V_{n,2})$, a homotopy
group of a Stiefel manifold. For a compact orientable seven-manifold this index
simply happens to vanish. This has been used in \cite{fkms} to show 
that\footnote{Note that the authors of \cite{fkms}
call a topological $G$-structure what we call a $G$-structure, and
a geometric $G$-structure what we call a torsion-free $G$-structure.
We also note that though the results of \cite{t} and \cite{fkms}
are stated for compact seven-manifolds, the proofs rely largely on the
dimensionalities and hold for non-compact case. } 
{\em a compact spin seven-manifold 
admits an SU(2)-structure.} As we said, this implies in particular 
SU(3) and G$_2$ structure. This simply means that instead of using both the
vectors we only use one or none respectively.

We now specialize as promised 
the above general discussion about $G$-structures and invariant tensors 
to the cases relevant to us. 

As mentioned above, all these $G$-structures for $G\subset$ SO(7) come
together with certain $G$-invariant spinors. 
The G$_2$ case is by now familiar: the invariant tensor which determines it 
is a three-form $\Phi$. This satisfies the octonionic structure constants:
\[
\Phi_{abe}\Phi^{cd}_{\ \ e}=2\delta_{ab}^{\ \ [cd]} - (*\Phi)_{ab}^{\ \ cd}\ .
\]
$\Phi$ singles out one G$_2$-invariant nowhere vanishing real spinor $\vth$,
in terms of which $\Phi_{abc}=-i\, \vth^\dagger \gamma_{abc}\vth$. 

An SU(3)-structure in seven dimensions is given by tensors $(v,J,\psi_3)$: 
a nowhere vanishing (which we take to be normalized to 1)
vector field $v$ (we will use the same symbol 
to denote the vector field and its dual (via the
metric) one-form), a generalized almost complex
structure $J$ (again the same symbol will denote its associated
two-form) and a three-form $\psi_3$. 
In order to define an
SU(3)-structure in seven dimensions they furthermore have to satisfy
\begin{subequations}
\begin{align}
(i) && v\, \lrcorner\, J &= 0 , \\
(ii) && v \, \lrcorner\, \psi_3 &= 0 , \\
(iii) && J^a_{~b}J^b_{~c} &= -\delta^a_{~c} + v^a v_c , \\
(iv) && \psi_3(X,JY,Z) &= \psi_3(JX,Y,Z) . 
\end{align}
\end{subequations}
These relations are loosely speaking a  ``dimensional reduction'' along $v$ of 
the octonionic structure constants given above. $\psi_3$ is to be thought of
as ${\rm Re}\{\Omega\}$.
We can determine the two- and three-form in terms of the vector and an
underlying G$_2$-structure as,
\beq
J= v \, \lrcorner \, \Phi
\quad \mbox{and} \quad
\psi_{3} = e^{2i\xi}
\left( \Phi - v\wedge (v\, \lrcorner\, \Phi) \right) 
 \label{J.psi} 
\eeq   
where the phase $e^{2i\xi}$ is a parameter of the SU(3)-structure. 

Again, besides such a description in terms of tensors there is one in terms
of spinors $\vth_\pm$, 
which is the one which comes naturally out of supersymmetry,
as outlined in the introduction. Again the tensors are bilinears of the 
spinor, as we will see in more detail in what follows. Conversely, given the
tensors one gets two SU(3)-invariant spinors on $Y$. 
The first one is
given by the G$_2$-invariant spinor $\vth$ associated to the
underlying G$_2$-structure on $Y$ and the second one by 
$v\cdot \vth\equiv v^a\ga_a \vth$.  
$\vth$ is usually taken to be of unit norm. 
Also $v \cdot \vth$ then has norm one. 

Now, if we take linear combinations  
$\vth_{+} = \left( 1+ v^a\ga_a \right)\vth$ as in (\ref{eq:dnw}), it is easy
to see that forming bilinears one gets back the tensors one started with. 
One might wonder whether this inverse is unique. 
A priori other linear combinations might work, but 
by Fierz identities one can show $v \vth_\pm = \pm \vth_\pm$, from which 
(\ref{eq:dnw}) follows. Interestingly, we will find the same result also in 
another way, while considering the differential equations coming from 
bilinears in the next section.

We finally also introduce intrinsic torsions, which we will later 
see practically into play.
This comes about while comparing
 connections on the bundles $TY$, $FY$ and $P$, the latter being the principal
bundle which defines the $G$-structure. 
Connections on $TY$ and $FY$ are  in one-to-one correspondence. 
Any connection on $P$ lifts to a unique connection on $FY$, 
whereas a connection on $FY$ reduces to a connection on the subbundle
$P$ if and only if its holonomy (and the holonomy of the corresponding
connection on $TY$) is contained in $G$. Hence connections on $P$
are in one-to-one correspondence with connections of holonomy $G$ on $TY$.
However, there can be an obstruction against finding any connections 
on $P$ that induce {\sl torsion-free} connections on $TY$. This obstruction
is called the intrinsic torsion of the $G$-structure $P$. 
If it is non-vanishing, then in particular the Levi-Civita connection
on $TY$ cannot have holonomy in $G$ and the 
normalized $G$-invariant tensors and
spinors are not covariantly constant in the Levi-Civita connection.
Its definition is given in terms of the torsion of the difference of any two
connections on $P$. All we need is that it is a section of 
$G^\perp\otimes TY$,
where $G^\perp$ is the quotient of $F$ by $adP$. Then we can decompose the 
tensor product in representations of $G$ and get a certain number of tensors 
which we can equally well call intrinsic torsion. 
The prettiest example for 
seven dimensions is given by G$_2$ intrinsic torsion. In this case 
$G^\perp$ is in the representation {\bf 7} of G$_2$, since the adjoint of
SO(7) decomposes under G$_2$ as {\bf 21}= {\bf 14} + {\bf 7}. Now 
we get G$_2^\perp\otimes TY$={\bf 7} $\otimes$ {\bf 7}=  {\bf 1}+{\bf 14}+
{\bf 27}+{\bf 7}. So what we will call intrinsic torsion are actually four
tensors $X_i$, $i=1,\ldots,4$ in these representations of G$_2$.

These objects are easier to calculate thanks to the following fact. For
general G$_2$-structure manifolds the invariant form $\Phi$ is not covariantly
constant, and so $\nabla \Phi$ gives another measure of how far one is from
having G$_2$ torsions; in fact it is the same as intrinsic torsion \cite{fg}.
In turn, all the information inside $\nabla \Phi$ are contained inside $d\phi$
and $d*\phi$. Decomposing these in G$_2$ representations gives us our $X_i$
as
\begin{equation}
  \label{eq:g2tors}
  d\Phi= X_1 *\Phi + X_4 \Phi + X_3\ , \qquad 
  d*\Phi= \frac43 X_4 *\Phi +X_2 \Phi\ .
\end{equation}
The first equation is a four-form and thus it contains {\bf 35}=
{\bf 27}+{\bf 7}+{\bf 1}; the second is a five-form and so it decomposes as 
{\bf 21}={\bf 14}+{\bf 7}. The {\bf 7} appear twice, but one can show that
it is actually the same tensor up to a factor, as shown in (\ref{eq:g2tors}).
A further way we mention to compute torsions, which we illustrate briefly 
only in this example, is directly through the spinor
equation. If one manages to put the right hand side of $D \vth=\ldots$ in the
form $K_{abc}q^{ab}\vth$ (using relations such as (\ref{g2.gamma.spin}), which 
we will explain later), $K_{abc}$ is already the torsion \cite{s,gp}.
In our case actually one can more efficiently put this right hand side in the
form $(q_a + q_{ab}\gamma^b)\vth$, again using relations (\ref{g2.gamma.spin}).
Here $q_{ab}$ can be a general tensor with two indices, which thus decomposes
as {\bf 1}+{\bf 14}+{\bf 27}+{\bf 7} again and thus contains again all 
the information about intrinsic torsion. The {\bf 7} gets also contributions
from $q_a$.

A similar story holds for SU(3) structures in six or seven dimensions. In six
one has torsions in ({\bf 3}+{$\bf\bar3$}+{\bf 1})\,$\otimes$ 
({\bf 3}+{$\bf\bar 3$}), 
which can be similarly as in the previous case encoded
in $dJ$ and $d\Omega$ \cite{gh,cs} (see also \cite{kmptt}). For seven 
dimensions the representations involved are even more, since we have
(2$\times${\bf 3}+2$\times${$\bf\bar 3$}+{\bf 1})\,$\otimes$ 
({\bf 3}+{$\bf\bar 3$}+{\bf 1}); again one can encode them in $dJ$, $d\Omega$ 
and $dv$, but this starts being of less practical use.

\section{Supersymmetry constraints}

The gravitino variation of eleven-dimensional supergravity in the
presence of a nontrivial 4-form flux $G=dC$ and with vanishing
gravitino background values reads 
\beq
\delta \hPsi_A = \left\{ \hD_A[\haom] +\frac{1}{144} \hG_{BCDE}
 \left(\hga ^{BCDE}_{~~~~~~~A} - 8 \hga ^{CDE} \eta ^B_{~A} \right) 
 \right\} \eps , \label{grav.flux.var.11}
\eeq
where $\eps$ is a Majorana spinor in eleven dimensions.
Our conventions on indices here and in the following are
\beq
\begin{array}{l@{~=~}l|l@{~=~}l}
\multicolumn{2}{c|}{\mbox{frame indices}} & 
\multicolumn{2}{c}{\mbox{coordinate indices}} \\ \hline
\rule{0mm}{4mm} A,B,\ldots & 1,\ldots,11 & 
M,N,\ldots & 1,\ldots,11 \\ 
\al,\beta,\ldots & 1,\ldots,4 &  
\mu,\nu,\ldots & 1,\ldots,4  \\ 
a,b,\ldots & 5,\ldots,11 &
m,n,\ldots & 5,\ldots,11  \\ 
\end{array} \non
\eeq
and hats refer to objects defined w.r.t. the eleven-dimensional
frame. The signature of eleven-dimensional spacetime is 
$(-++\ldots +)$ and the $\hga$-matrices satisfy 
$\{\hga_A,\hga_B\}=2\eta_{AB}\ident$.

We want to consider warped compactifications 
\beq
d\hat{s}_{11}^2 = e^{2\Delta} ds_4^2 + ds_7^2 , \label{ds11.w}
\eeq
where the warp factor depends only on the internal coordinates, 
$\Delta=\Delta(x^m)$ and where the four-dimensional spacetime 
with metric $ds_4^2$ is
Minkowski.
Lorentz invariance requires the background flux to be of the form
\begin{align}
G &= 3 \mu \frac{1}{4!} 
\eps_{\mu\nu\rho\sigma} dx^{\mu\nu\rho\sigma} + \frac{1}{4!}
G_{mnpq} dx^{mnpq} \non \\
&= 3 \mu \frac{1}{4!} 
\eps_{\al\beta\ga\delta} e^{\al\beta\ga\delta} + \frac{1}{4!}
G_{abcd} e^{abcd} 
 \label{ansatz.G}  \\
&= 3 \mu e^{-4\Delta} \frac{1}{4!} 
\eps_{\al\beta\ga\delta} \he^{\al\beta\ga\delta} + \frac{1}{4!}
G_{abcd} \he^{abcd} \non 
\end{align}
with a real constant $\mu$. 

The decomposition for the $\hga$-matrices is the standard one,
\begin{subequations}
\begin{align}
\hga^{\al} &= \tga^{\al} \otimes \ident & \quad & \mbox{for}~\al=1,\ldots
,4 \\
\hga^a &= \tga^{(5)} \otimes \ga^a & \quad & \mbox{for}~a=5,\ldots
,11 
\end{align}
\end{subequations}
where $\tga^{(5)}=i\tga^1\tga^2\tga^3\tga^4$ is the four-dimensional
chirality operator. For explicit computations we will use the Majorana
representation in which the
$\ga$-matrices are either real ($\tga^{\al}$) or imaginary
($\tga^{(5)}$ and $\ga^a$). In this representation
the Majorana condition $\eps$ reduces to a reality constraint,
$\eps^*=\eps$. Using now projectors $P_\pm \otimes \Bbb I$ we can write now 
our class of $\eps$ as 
\beq
\eps=\psi_+ \otimes \vth_+ + \psi_- \otimes \vth_-  \label{ans.eps}
\eeq
again by four-dimensional invariance; also, 
$\vth_{\pm}$ depend only on the internal coordinates
$x^m$. The four-dimensional spinors 
$\psi_{\pm}$ are (covariantly) constant
\[ 
D_{\al}\psi_{\pm} = 0
\]
and chiral, $\tga^{(5)}\psi_{\pm} = \pm \psi_{\pm}$. The Majorana 
constraint on $\eps$ then requires $(\psi_{\pm})^*=\psi_{\mp}$ and 
$(\vth_{\pm})^*=\vth_{\mp}$. 
The gravitino variations 
(\ref{grav.flux.var.11}) then lead to the following supersymmetry
constraints on the internal spinors,
\begin{subequations}
\begin{align}
0&= \left[ \pm \left( -\mu i e^{-\Delta}  
 + \, \frac{1}{2} (\partial_c \Delta) \ga^c \right)
 + \frac{1}{144} 
  G_{bcde} \ga ^{bcde} \right] \vth_{\pm} \label{var.const.1}\\
\intertext{from the spacetime part $\al=1,\ldots,4$ and}
D_a[\om] \vth_{\pm} &= 
 \left[ 
 \mp \frac{1}{144}
 \left( G_{bcde} \ga ^{bcde}_{~~~~~a} - 8 G_{abcd} \ga ^{bcd}
   \right) \right] \vth_{\pm} , \label{var.const.2} \\ 
&=  \left[ 
 \pm  \left( \frac{i}{12} (*G)_{abc} \ga^{bc} 
 + \frac{1}{18} G_{abcd} \ga ^{bcd}
   \right) \right] \vth_{\pm}  \label{var.const.3}
\end{align}
\end{subequations}
from the internal part $a=5,\ldots,11$, 
where we have 
defined $(*G)_{abc}\equiv \frac{1}{4!}\eps_{abcdefg}G^{defg}$.

\subsection{Four-dimensional supersymmetry}

A method to get equations from (\ref{var.const.1})
 is simply to consider bilinear expressions 
\begin{equation}
  \label{eq:dilatino}
\vth_\pm \{G_{abcd}\gamma^{abcd}, \gamma_{a_1\ldots a_k}\}\vth_+  
\end{equation}
(and the same with $\{ \, , \, \} \rightarrow [ \, , \, ]$) and use
(\ref{var.const.1}).  
It might a priori be non obvious which and how many of them generate all the 
possible relations. The spinor representation in seven dimensions however
decomposes as 
{\bf 8}$\to${\bf 7}+{\bf 1}$\to${\bf 3}+${\bf \bar 3}$+2$\times${\bf 1}, 
which are $\gamma^a \vth_\pm$ and $\vth_\pm$. This simple
fact suggests the answer: generating relations come from (\ref{eq:dilatino})
for $k=0,1$. An equivalent but maybe more informative way of putting this ---
and which does not make use of the bilinears, whose structure we will determine
later --- is as follows.

The fact that the G$_2$ spinor is invariant implies, via the usual 
infinitesimal transformation for spinors $\delta \vth\sim q_{ab}
\gamma^{ab}\vth$,
that $\gamma^{ab}\vth$ belongs to the {\bf 7}. Using the bilinear expression 
for $\Phi$ allows one to fix the constants as 
\[
\ga_{ab} \vth = ~i \Phi_{abc} \ga^c \vth  
\]
a relation well-known in the context of manifolds of G$_2$ {\sl holonomy}, and
which is in fact valid in general for manifolds of G$_2$ {\sl structure}.
Group theory then allows one to determine for every 
$\gamma^{a_1\ldots a_k}\vth$ which representations are present and which ones
are not. For example, for $k=3$ we have that three-forms contain both {\bf 7}
and {\bf 1}, while for $k=5$ one only has {\bf 7}. Coefficients can then be
fixed by gamma matrix algebra and/or dualization. All this results in the
relations
\begin{align}
\ga_a \vth ~~~ & {}  & \in & \wedge^1_7 , \non \\
\ga_{ab} \vth = & ~i \Phi_{abc} \ga^c \vth  & \in & \wedge^2_7 , \non \\
\ga_{abc} \vth = & ~i \Phi_{abc} \vth - (*\Phi)_{abcd} \ga^d \vth 
 & \in & \wedge^3_1 \oplus \wedge^3_7 
, \label{g2.gamma.spin} \\ 
\ga_{abcd} \vth = & ~(*\Phi)_{abcd} \vth - 4i \Phi_{[abc} \ga_{d]} \vth &
 \in & \wedge^4_1 \oplus \wedge^4_7 , \non \\
\ga_{abcde} \vth = & ~5 (*\Phi)_{[abcd} \ga_{e]} \vth &
 \in & \wedge^5_7 , \non \\
\ga_{abcdef} \vth = & ~i \eps_{abcdefg} \ga^{g} \vth &
 \in & \wedge^6_7 . \non
\end{align}
Here $\wedge^k_l$ denotes the irreducible G$_2$-representation of dimension
$l$ in the decomposition of $\wedge^k T^*Y$. 
Furthermore $\Phi_{abc}$ are the
components of the G$_2$-invariant three-form $\Phi$ and are given by
the structure constants of the imaginary octonions. When taking bilinears,
most of the terms in (\ref{g2.gamma.spin}) drop out due to
$\vth^{\dagger}\ga_a
\vth=0$ leaving only the familiar $\Phi_{abc} =
-i\vth^{\dagger}\ga_{abc}\vth$ and $(*\Phi)_{abcd} =
\vth^{\dagger}\ga_{abcd}\vth$. We do emphasize however the importance of
terms in $\wedge^3_7$ and $\wedge^4_7$ for our analysis.
Similar but less pretty relations can be obtained
for the SU(3) structure directly. In what follows it turns 
out sufficient to
use just (\ref{g2.gamma.spin}) and the explicit expression 
for the spinors $\vth_\pm$ (see \ref{ans.vthpm.2})).

We define the following projections of the four-form flux
\begin{equation}
Q \equiv \frac{1}{4!} G_{abcd} (*\Phi)^{abcd}\ , \qquad
Q_a \equiv \frac{1}{3!} G_{abcd} \Phi^{bcd} \ , \qquad
Q_{ae} \equiv \frac{1}{3!} G_{abcd} (*\Phi)^{bcd}_{~~~e}\ ; 
\end{equation}
equivalently we can write 
\begin{equation}
  \label{eq:Grep}
  G_{abcd}= \frac{4}{7}Q (*\Phi)_{abcd}+  Q_{[a}
\Phi_{bcd]}- 2 
\hat Q_{e[a} (*\Phi)^e_{\ bcd]}\ .
\end{equation}
$Q$ and $Q_a$ only contain the projections onto the singlet and the
{\bf 7} in the decomposition of the internal four-form flux
into G$_2$-representations. $\hat Q_{ab}\equiv Q^{\bf 27}_{ab}$ is 
symmetric and traceless. We note that all contractions of $G$ with
$\Phi$ and $*\Phi$ can be expressed solely in terms of $Q$, $Q_a$ and
$\hat
Q_{ab}$. In particular, $Q_{ab} = -\frac{4}{7} Q \delta_{ab} + \frac12
\Phi_{abc}Q^c + \hat Q_{ab}$.
Plugging (\ref{g2.gamma.spin}) into (\ref{var.const.1}) and using the linear
independence of $\vth$ and $\ga^a\vth$, we obtain expressions of the form
$(A+B_a \gamma^a)\vth$. Thus we see explicitly that $A$ and $B_a$ have to be
put to zero and that this is all the information in the four dimensional 
equation. These are, writing real and imaginary parts separately,
\begin{subequations}
\begin{align}
\mu&=0 \quad , \quad
& Q_a &=3 \Phi_{abc}v^b\partial^c \Delta, \label{eq:dila}\\
Q_{ab}v^a v^b&=0 \quad , \quad
& -3 \partial_a \Delta& = Q v_a + Q_{\{ab\}}v^b \ \label{eq:dila2} .
\end{align}
\end{subequations}
This in particular sets to zero the Freund-Rubin parameter, which is
not surprising as we are on Minkowski, and gives relations between $G$ and
$d\Delta$ on which we already commented in the introduction. We will find
again some of these in a maybe more palatable form.
From (\ref{eq:dila}) one can also derive 
\begin{equation}
  \label{nogo}
9(\partial_a\Delta\partial^a\Delta)=Q_a Q^a + Q^2   \ ,
\end{equation}
from which one can see that in particular there cannot be any warped 
compactification if $G$ has only components in the {\bf 27}. 
On the other side,
(\ref{eq:dila2}) implies that it's impossible to have $G$ only in the 
{\bf 1} or only in the {\bf 7} either, since these are determined by  
$\hat Q_{ab}$. We recall \cite{cr} that when the warp factor is absent, all 
components of flux must vanish. So far we see that taking $\Delta$ to be 
constant
gives $Q=Q_a = \hat Q_{ab} v^a = 0$. The rest of the conditions, like in \cite{cr} 
should come from analyzing the internal part.
We will return to these points shortly in subsection
\ref{short}.

\subsection{Internal part}

In order to study the constraints that (\ref{var.const.2}) impose on
the geometry, we introduce the following bilinears,
\beq
\Xi_{a_1\ldots a_n} \equiv
 (\vth_+)^{\dagger} \ga_{a_1 \ldots a_n} \vth_+ 
\qquad \mbox{and} \qquad
\tilde\Xi_{a_1\ldots a_n} \equiv
 (\vth_-)^{\dagger} \ga_{a_1 \ldots a_n} \vth_+\ 
 \label{def.omtom} 
\eeq
and their associated forms
\beq
\Xi_n \equiv \frac{1}{n!} \Xi_{a_1\ldots a_n} e^{a_1\ldots a_n} 
\qquad \mbox{and} \qquad
\tilde\Xi_n \equiv \frac{1}{n!} \tilde\Xi_{a_1\ldots a_n} e^{a_1\ldots a_n}\ .
\eeq

The full set of bilinears for $n=1,...,7$ is obviously redundant. One way
to see relations between them would be to
use Fierz identities. A faster way in this case is to use the expression
for $\vth_\pm$ in terms of $v$ and $\vth$. But before we want to fix again
possible ambiguities in that expression. We want to ask what are the 
normalizations $\vth_+^\dagger\vth_+=\vth_-^\dagger\vth_-=\Xi$ and the
scalar product $\vth_-^\dagger\vth_+=\tilde\Xi$ 
(we have dropped the subscript $0$ on the functions). For these we can derive 
differential equation using the usual methods, based on
the supersymmetry constraints (\ref{var.const.1}) and
(\ref{var.const.2}):
\begin{equation}
  d\left( e^{-\Delta} \Xi \right) = 0 , \quad \quad \quad
d\left( e^{2\Delta} \tilde\Xi \right) = 0 , \label{dOmdtOm} \ .
\end{equation}
To these one has to add another remark \cite{gmw}.
Note that since $\vth_{\pm}$ are by construction invariant
spinors of an SU(3)-structure and $(\vth_{\pm})^*=\vth_{\mp}$, there
exists a connection w.r.t.~which the normalized spinors
$
\frac{1}{\sqrt{\Xi}} \vth_{\pm}$ are covariantly
constant. This implies that the scalar product between the normalized spinors
\beq
\frac{\vth_-^\dagger\vth_+}
{\sqrt{\vth_-^\dagger\vth_-}\sqrt{\vth_+^\dagger\vth_+}}=
\frac{\tilde\Xi}{\Xi} 
\eeq
has to be constant. For nontrivial warp factor $\Delta(x^m)$ this,
combined with
(\ref{dOmdtOm}), forces $\tilde\Xi$ to vanish, so that the spinors have
to be orthogonal and we find (\ref{eq:dnw}) again. Moreover, we can also 
substitute there $\Xi= e^\Delta$ for the normalization.
We write again the result for $\vth_\pm$:
\beq
\vth_{\pm} = \frac{1}{\sqrt{2}} e^{\frac{1}{2}\Delta\pm i \xi} 
 \left( \ident \pm v^a\ga_a \right) \vth  \ .\label{ans.vthpm.2}
\eeq
Note that the phase $\xi$ reflects the existence of U(1) of SU(3)
structures inside G$_2$ (see \ref{J.psi}).

Using this we can now compute easily
\begin{subequations}
\begin{align}
\Xi_{\rule{1.5mm}{0mm}} &= e^{\Delta} , \label{fierz}\\
\Xi_1 &= e^{\Delta}v , \\
\Xi_{2} &=  i e^{\Delta} v\, \lrcorner\, \Phi , \\
\Xi_{3} &= i e^{\Delta} v \wedge (v\, \lrcorner\, \Phi)
 = e^{-\Delta} \Xi_1 \wedge \Xi_2 , \\
\tilde\Xi_{3} &= e^{\Delta+2i\xi} \left[
 i \Phi -i  v \wedge (v\, \lrcorner\, \Phi)
 - v\, \lrcorner \, (*\Phi) \right]  \\ 
&= e^{\Delta+2i\xi} \left[
 i \Phi - e^{-2\Delta} \Xi_1 \wedge \Xi_2 
 - v\, \lrcorner \, (*\Phi) \right] , \\ 
\Xi_{4} &= e^{\Delta} \left[
 (*\Phi) -  v\wedge (v\, \lrcorner\, (*\Phi)) \right] , \\
\tilde\Xi_{4} &= e^{\Delta+2i\xi} \left[
 v\wedge (v\, \lrcorner (*\Phi)) -i v\wedge \Phi \right] 
 = -e^{-\Delta} \Xi_1 \wedge \tilde\Xi_3 \ ;
\end{align}
\end{subequations} 
the properties of these bilinears, as for example the vanishing of 
$\tilde\Xi_i$ for $i=1,2$ and their duals, are consequence of properties of
gamma matrices in seven dimensions.
We see that a basis of generators are $\Xi_1$, $\Xi_2$ 
and $\tilde\Xi_3$. These are, up to normalization factors $e^\Delta$, 
the invariant tensors which characterize our SU(3) structure in seven 
dimensions. What is now left is to compute differential equations for these
tensors using again the supersymmetry constraints (\ref{var.const.1}) and
(\ref{var.const.2}). We actually do better and compute them for all the 
tensors we wrote:
\begin{subequations}
\begin{align}
d\left( e^{\Delta} \Xi_1 \right) &
=0\ , \quad 
& e^{-3\Delta} d\left( e^{3\Delta} \Xi_2 \right) &= 
 -2i\, \Xi (*G) \label{diffeq}
\end{align}
and
\begin{align}
e^{-5\Delta} d\left( e^{5\Delta} \Xi_3 \right) 
 &=  
2i\, \Xi_1 \wedge (*G)\ , \\ 
e^{-2\Delta} d\left( e^{2\Delta} \tilde\Xi_3 \right) 
 &= 
-2 \tilde\Xi G\ , \\
e^{-\Delta} d\left( e^{\Delta} \Xi_4 \right) 
  &=  3  G \wedge \Xi_1 \ , \\
e^{-4\Delta} d\left( e^{4\Delta} \tilde\Xi_4 \right) 
 &= 0\ .\label{diffeq.f}
\end{align}
\end{subequations} 

As already pointed out the set of $\Xi_n$ and $\tilde\Xi_n$ is redundant
and thus
the systems (\ref{diffeq}) and (\ref{fierz}) lead to extra consistency
conditions. One might wonder whether these are new constraints to be added
to  (\ref{eq:dila}). Due to the following argument, this can only be the
case if we get relations involving $G$.  A priori, one could have computed
differential equations for (\ref{diffeq} -- \ref{diffeq.f}) using only
(\ref{var.const.2}) and not (\ref{var.const.1}). The result is a
collection of rather cumbersome  expressions involving partial
contractions of $G$ with $\Phi$ and $*\Phi$ (and no  $d \Delta$).
Thus any
extra constraint in consistency  conditions among (\ref{diffeq} --
\ref{diffeq.f}), must be expressible in a form involving $G$ only. This is
not the case: the checks of consistency of (\ref{diffeq}) with
(\ref{fierz}) yield the single equation
\begin{equation}
  \label{eq:mono}
  6 d\Delta\wedge \Xi_4 = -2i (*G) \wedge \Xi_2 -3 \,G \wedge\Xi_1\ .
\end{equation}
Being a five-form, this equation can be split into ${\bf 7}$ and ${\bf 14}$ 
components. The vector part contains $d\Delta$, 
and thus it must be dependent on the conditions coming from 
(\ref{var.const.1}).
The part in the ${\bf 14}$ is instead purely in terms of $G$ and is 
independent. 
One might want to consider this as a kind of ``monopole equation'' for the  
present class of compactifications. One can prove the following relations
for a normalized vector $v$ that are particularly useful in the checks
mentioned above:
\begin{subequations}
\begin{align}
v \wedge (v \lrcorner \Phi) \wedge (v \lrcorner \Phi) &= (v \lrcorner
\Phi) \wedge \Phi, \\
(v \lrcorner \Phi) \wedge (v \lrcorner *\Phi) &= 0 \\
(v \lrcorner \Phi) \wedge (v \lrcorner \Phi) &= - 2\left(*\Phi - v \wedge
(v \lrcorner (*\Phi))\right).
\label{consixt}
\end{align}
\end{subequations} 
The last relation implies in particular $\Xi_4 = \frac{e^{-\Delta}}{2} 
\Xi_2 \wedge \Xi_2$.

\subsection{Intrinsic torsion}

Note that the two- and three-form defining the SU(3)-structure
(\ref{J.psi}) are given in terms of $\Xi$'s as
\beq 
J= -i e^{-\Delta} \Xi_{2} \qquad \mbox{and}\qquad
\psi_3=e^{2i\xi} \im(e^{-\Delta-2i\xi} \tilde\Xi_{3}) ,
\eeq 
whereas the three- and four-form of the G$_2$-structure have
representations, 
\begin{align}
\Phi &= \im(e^{-\Delta-2i\xi} \tilde\Xi_{3})-ie^{-2\Delta} \Xi_1 \wedge
\Xi_2 , \\
*\Phi &= e^{-\Delta} \Xi_4 -e^{-2\Delta} \Xi_1\wedge 
 \re(e^{-2i\xi}\tilde\Xi_3) .  
\end{align}
The equations (\ref{eq:de}) were written in terms of the maybe more familiar
complex three-form $\Om$, which in terms of the above reads 
$\Omega\equiv \psi_3 +iv\lrcorner(*\psi_3)$.

We can finally compute G$_2$ intrinsic torsions as promised. For this we need  
\begin{subequations} 
\begin{align}
d\Phi &= -3d\Delta \wedge [ \Phi+v\wedge (v\, \lrcorner \, \Phi )]
 + 2d\xi \wedge [ v \, \lrcorner \, (*\Phi)]
 + 2 v \wedge (*G) , \\
d(*\Phi) &= -2 d\Delta \wedge (*\Phi) 
-3 d\Delta \wedge v \wedge [ v \, \lrcorner \, (*\Phi)]
 -2d\xi \wedge v\wedge  \Phi  
 + 3 G\wedge v 
\end{align}
\end{subequations}
Projecting these into representations we get (up to overall factors)
\begin{subequations} 
\begin{align}
X_{\bf 1}&=\partial_a \xi v^a \ , \label{x1}\\
(X_{\bf 14})_{ab}&=\frac12 \Phi_{abef}v^e Q^f+2v_{[a}Q_{b]} 
-\frac12 \Phi_{abc}\hat Q^c_{\ e}v^e + \hat Q^c_{\ [a}\Phi_{b]ec}v^e \ ,
\\
(X_{\bf 7})_a& =- 8\, \Phi_{abc}v^b\partial^c\xi +3 \hat Q_{ab}v^b 
-15 \partial_a \Delta + \frac{65}7 Q v_a \ , \\
 (X_{\bf 27})_{ab}=& -4 \left(\partial_{\{a}\xi
v_{b\}}-\frac17\delta_{ab}(\partial_c\xi v^c)\right) - Q_{\{a}v_{b\}} +
\hat Q_{\{a}^{\ \ e} \Phi_{b\}ec}v^c\ 
\label{x27}.
\end{align}
\end{subequations} 
Here we denoted torsions by representations. Thanks to 
(\ref{eq:dila},\ref{eq:dila2}), one 
can also derive
\begin{equation}
  \label{eq:final}
 v_a= \frac1{\partial_e \Delta \partial^e \Delta}
\left[ 6 \Phi_{abc}Q^b\partial^c\Delta -\frac Q3 \partial_a \Delta\right]
\end{equation}
to eliminate $v$ and make (\ref{x1},\ref{x27}) purely in terms of physical
quantities. Note also that in all these expressions $Q$ and $Q_a$ can also
be eliminated in favor of $\hat Q_{ab}$.

Of course in a sense these expressions do not mean that one can 
forget about $v$ altogether:
one still has to check the differential equation for $v$, 
\begin{equation}
  \label{eq:dv}
d(e^{2\Delta} v)=0  \ ,
\end{equation}
separately. But we have decoupled $dv$ from $d\Phi$. About (\ref{eq:dv}) we
can actually say more: it also has a mathematical meaning.
It is known that this implies \cite{s,gmw,gmpw} that the 
seven-dimensional metric can be written in a product form
\[
ds^2_7=ds^2_6 +e^{-4\Delta}dx^2_7
\]
with no restriction, however, on the coordinate dependence of $\Delta$
and $ds^2_6$. Indeed $v$ 
needs not be Killing: the symmetric part of its covariant derivative, which
we have not written above, reads 
\begin{equation}
  \label{eq:killingl}
 D_{\{a} \Xi_{b\}}= \frac{1}{3} Q \delta_{ab} -  Q_{\{ab\}}\ .
\end{equation}
So if we want this to be a Killing vector, we have to impose that $G$ is in
the {\bf 7} only. But, as already noticed, 
from (\ref{eq:dila2}) one sees that if $Q$ and
$Q_{\{ab\}}$ vanish, the warp factor $\Delta$ is constant.  In fact it is not hard to see that the equation (\ref{eq:killingl}) implies  the second equation in (\ref{eq:dila2}).

\subsection{A short summary}
\label{short}

We have presented here a set of general relations between the components
of intrinsic torsion on a generic seven-dimensional manifold, admitting
spinors and thus a G$_2$ structure, and the components of four-form flux.
Due to the existence of a full classification of manifolds admitting G$_2$
structure, one may hope that a similar
classification of M-theory backgrounds can be achieved. 
We would like to emphasize that
the results presented here are just a set of necessary conditions for
preserving supersymmetry.  Namely given a manifold with a particular set
of intrinsic torsions, we know now what is the possible profile of the
four-form flux needed for preserving supersymmetry, and vice versa.  
While the existence of the SU(3) structure is crucial for preserving
${\cal N}=1$ supersymmetry, as explained above on {\it any}
seven-dimensional spin manifold this structure is already present due to
existence of (two!) nowhere vanishing vector field(s). Supersymmetry does
however impose a differential equation on this vector field (an analogue
of the Killing vector equation). A general analysis of existence of
solutions for this equation might be an interesting problem, which is
beyond the scope of this paper.

It is time now to collect all the information concerning the four-form flux.
While the relations between the components of the flux and intrinsic
torsion is in general complicated, we see that supersymmetry imposes
strong constraints on components of the flux. In particular, the
``primitive" part, namely $G_{\bf 27}$ is the most important part of
the flux, and determines the two other components, through the
expressions
\begin{equation}
  \label{eq:whocares}
  Q= \frac74 \hat Q_{ab}v^a v^b\ , \qquad Q_a= -2 \Phi_{abc}v^b \hat
  Q^c_{\ d}v^d\ .
\end{equation}
It is easy to see that vanishing
of {\bf 27} leads to vanishing of the other two components, but not
the other way around. Thus, $G_{\bf 27}$ 
cannot be zero. From other side, due to (\ref{nogo}), it cannot be the
only component turned on. Thus in order to have a warped compactification, 
primitivity is not enough.  

Going back to the case of constant warp factor, one can see that
to the conditions  $Q=Q_a = \hat Q_{ab} v^a = 0$, stated above, equation (\ref{eq:mono}) 
adds $ \hat Q^c_{\ [a}\Phi_{b]ec}v^e = 0$. This still does not eliminate 
$\hat Q_{ab}$, and thus $G_{\bf 27}$, entirely, and one has to go back to the 
integrability conditions \cite{cr}.  For $\Delta =0$ case, the integrability is certainly
the most restrictive, since the 
Ricci scalar is negative semi-definite and the equation 
$-R + G^2 = 0$ forces both the vanishing of flux and Ricci-flateness.  This condition 
is much less restrictive (and less useful) for a non-constant warp factor, since now the 
Ricci scalar is no longer  semi-definite  and the equation acquires new terms like
$\Phi^{abc} D_aX_{bc}$ which are not positive-definite. In other words, for
a warped product involving  a 
generic seven-manifold with components of intrinsic torsion $X$, after having built 
$G_{\bf 1}$ and $G_{\bf 7}$ in terms of the primitive flux according to 
(\ref{eq:whocares}), one has enough  freedom to satisfy (\ref{eq:mono}) for a non-trivial
four-form  $G_{\bf 27}$.

\section{SU(2) structures and towards $\cN=2$.}

We already saw that the existence on the internal manifold of vector fields
without zeros can have rather far-reaching consequences for supersymmetry.
We have concentrated so far  on the ``minimal" case of SU(3) structure 
with ${\cal N}=1$ supersymmetry where only one such vector was actively
involved.  As we saw the SU(3) structure comes out rather 
naturally from the seven-dimensional parts $\vth_\pm$ of
$\frac12 (1\pm \gamma^{(5)})\otimes\Bbb I \epsilon$.

However since we have a pair of vector fields on seven-manifolds and
thus, as discussed in section \ref{prelim}, SU(2) structure, it is natural
to ask what the consequences of this are for supersymmetry. Of course, one
could easily add the second  vector field in (\ref{ans.vthpm.2}) by $v
\rightarrow \sum a_i v^i$, but it is not hard to see that this replacement
preserves as much supersymmetry as the original Ansatz. Moreover since we
have argued that (\ref{ans.vthpm.2}) is the most general possibility, by
field redefinitions we can bring the new spinor to this form again. To be
short, using SU(2) structure to preserve $\cN=1$ supersymmetry does not
give anything new in comparison to SU(3) case.

The situation will be different when one will want to look for $\cN=2$
solutions where one can use the SU(2) structure in a more interesting
fashion. Here we make the first step in that direction by writing down the
generalization of (\ref{ans.vthpm.2}) suitable for four-dimensional
$\cN=2$. 

Strictly speaking one would only need three spinors to define an SU(2) 
structure. But we can easily come up with a fourth one, which does not add
extra structure, but is more compatible with four-dimensional chirality. 
We will now have four spinors $\vth_\pm^i$, $i=1,2$.
The three spinors of SU(2) structure can be thought of as
$\vth$, $v_1^a \gamma_a \vth$ and $v_2^a \gamma_a \vth$, where we now make
use of both the vectors we discussed in section \ref{prelim}. Then the fourth
spinor can be easily constructed as the Clifford action of both $v_i$ on 
$\vth$, that is 
$\tilde\vth\equiv v_1^a v_2^b \gamma^{ab} \vth$. If one wants, this too can be
cast in a form similar to the other spinors writing it as 
$\tilde\vth=v_3^a \gamma_a \vth$, where 
$v_3^a\equiv i\Phi^a_{\ bc}v_2^b v_3^c$.

We can now combine these four spinors in a way similar to (\ref{ans.vthpm.2}),
to produce the four spinors we want $\vth_\pm^i$ ($i=1,2$). We still want
them to be 
orthogonal to each other for arguments similar to those leading to 
(\ref{ans.vthpm.2}); the U(1) freedom that we had before 
(the phase $e^{i\xi}$) gets now replaced by a U(2) freedom. So one can start
from whatever choice and act with a U(2) matrix with it. A possibility to
express this is, very explicitly,
\begin{subequations}
  \begin{align}
 \label{eq:n2}
  \vth_+^1=& e^{i\xi} (a + b v_1 + a v_2 + b v_3)\cdot \vth\\
  \vth_+^2=& e^{i\xi} (-\bar b+ \bar a v_1 -\bar b v_2 +\bar a v_3)\cdot
\vth
\end{align}
\end{subequations}
where we have now denoted the Clifford multiplication by a dot, 
$v\cdot\vth\equiv v_a \gamma^a\vth$; the $\vth^i_-$ are then 
$\vth_-^i=(\vth_+^i)^*$, and one has to remember that
$v_3$ is purely imaginary.
The eleven-dimensional spinor now is $\eps=\psi^i_+ \otimes \vth^i_+ +
\psi^i_- \otimes \vth^i_- $.

Thus indeed the existence of SU(2) structure on seven-manifolds
leads  to possibility of preserving ${\cal N}=2$ supersymmetry. The
possibility of starting with a certain background
and enhancing supersymmetry by adjusting the fluxes looks interesting and
to our opinion deserves further study.

\bigskip\bigskip

{\bf Acknowledgments.} We would like to thank  Michela Petrini for
collaboration in initial stages of this work. We also thank Mariana
Gra\~na, Dario Martelli, Andrei Moroianu, Stefan Theisen and Daniel
Waldram for many useful conversations.
This work is supported in part by EU contract HPRN-CT-2000-00122 and by
INTAS contracts 55-1-590 and 00-0334. PK is supported by a
European Commission Marie
Curie Postdoctoral Fellowship under contract number
HPMF-CT-2000-00919, AT is supported by IPDE.



\begin{thebibliography}{99}

\bibitem{gmpw}J.~P.~Gauntlett, D.~Martelli, S.~Pakis and D.~Waldram,
``G-structures and wrapped NS5-branes,'' arXiv:hep-th/0205050. 

\bibitem{gmw}J.~P.~Gauntlett, D.~Martelli and D.~Waldram,
``Superstrings with intrinsic torsion,'' arXiv:hep-th/0302158. 

\bibitem{glmw}S.~Gurrieri, J.~Louis, A.~Micu and D.~Waldram,
``Mirror symmetry in generalized Calabi-Yau compactifications,'' Nucl.\
Phys.\ B {\bf 654}, 61 (2003) [arXiv:hep-th/0211102]. 

\bibitem{mgH} G.~L.~Cardoso, G.~Curio, G.~Dall'Agata, D.~Lust, P.~Manousselis 
and G.~Zoupanos,
``Non-Kaehler string backgrounds and their five torsion classes,'' Nucl.\
Phys.\ B {\bf 652}, 5 (2003) [arXiv:hep-th/0211118]. 

\bibitem{kmpt}P.~Kaste, R.~Minasian, M.~Petrini and A.~Tomasiello,
``Kaluza-Klein bundles and manifolds of exceptional holonomy,'' JHEP {\bf
0209}, 033 (2002) [arXiv:hep-th/0206213]. 

\bibitem{kmptt}P.~Kaste, R.~Minasian, M.~Petrini and A.~Tomasiello,
``Nontrivial RR two-form field strength and SU(3)-structure,''
arXiv:hep-th/0301063. 

\bibitem{gp}J.~P.~Gauntlett and S.~Pakis,
``The geometry of D = 11 Killing spinors,'' arXiv:hep-th/0212008.

\bibitem{cr} P.~Candelas and D.~J.~Raine,
``Spontaneous Compactification And Supersymmetry In D = 11 Supergravity,''
Nucl.\ Phys.\ B {\bf 248}, 415 (1984).

\bibitem{dnw}
B.~de Wit, H.~Nicolai and N.~P.~Warner,
``The Embedding Of Gauged N=8 Supergravity Into D = 11 Supergravity,''
Nucl.\ Phys.\ B {\bf 255} (1985) 29.

\bibitem{as}  B.~S.~Acharya and B.~Spence, ``Flux, supersymmetry and M theory
on 7-manifolds,'' arXiv:hep-th/0007213. 
 
\bibitem{bb7} K.~Becker and M.~Becker, ``Compactifying M-theory to four
dimensions,'' JHEP {\bf 0011}, 029 (2000) [arXiv:hep-th/0010282].

\bibitem{ali} T.~Ali, ``M-theory on seven manifolds with G-fluxes,''
arXiv:hep-th/0111220. 

\bibitem{bj} K.~Behrndt and C.~Jeschek, ``Fluxes in M-theory on 7-manifolds
and G structures,'' arXiv:hep-th/0302047. 


\bibitem{fs} B.~Brinne, A.~Fayyazuddin, S.~Mukhopadhyay and D.~J.~Smith,
``Supergravity M5-branes wrapped on Riemann surfaces and their QFT
duals,'' JHEP {\bf 0012}, 013 (2000) [arXiv:hep-th/0009047]; 
B.~Brinne, A.~Fayyazuddin, T.~Z.~Husain and D.~J.~Smith, 
``N =1 M5-brane geometries,'' JHEP {\bf 0103}, 052 (2001)
[arXiv:hep-th/0012194]. 

\bibitem{cco} M.~Cvetic, H.~Lu and C.~N.~Pope, ``Massless 3-branes in
M-theory,'' Nucl.\ Phys.\ B {\bf 613}, 167 (2001) [arXiv:hep-th/0105096];
M.~Cvetic, G.~W.~Gibbons, H.~Lu and C.~N.~Pope, ``M3-branes,
G(2) manifolds and pseudo-supersymmetry,'' Nucl.\ Phys.\ B {\bf 620}, 3
(2002) [arXiv:hep-th/0106026]. 

\bibitem{mt} R.~Minasian and D.~Tsimpis, ``Hopf reductions, fluxes and
branes,'' Nucl.\ Phys.\ B {\bf 613}, 127 (2001) [arXiv:hep-th/0106266].



\bibitem{mn}J.~M.~Maldacena and C.~Nunez, ``Supergravity description of
field theories on curved manifolds and a no go theorem,'' Int.\ J.\ Mod.\
Phys.\ A {\bf 16}, 822 (2001) [arXiv:hep-th/0007018]. 

\bibitem{dhs}B.~de Wit, D.~J.~Smit and N.~D.~Hari Dass,
``Residual Supersymmetry Of Compactified D = 10 Supergravity,''
Nucl.\ Phys.\ B {\bf 283} (1987) 165.


\bibitem{fkms}Th.~Friedrich, I.~Kath, A.~Moroianu and U.~Semmelmann, 
``On nearly parallel G$_2$-structures,'' J.\ Geom.\ Phys.\ 
{\bf 23} (1997), 259-286.


\bibitem{bb8}K.~Becker and M.~Becker, ``M-Theory on Eight-Manifolds,''
Nucl.\ Phys.\ B {\bf 477}, 155 (1996) [arXiv:hep-th/9605053].


\bibitem{j}D.~Joyce, {\it Compact manifolds with Special Holonomy}, Oxford, 
2000.
\bibitem{t}E.Thomas, ``Vector fields on manifolds,'' Bull.\ Amer.\ Math.\ 
Soc.\ {\bf 75} (1969),643-683.
\bibitem{fg}M.~Fernandez and A.~Gray, ``Riemannian manifolds with structure
group G$_2$,'' Ann.\ di Mat.\ Pura ed Appl.\ {\bf 32} (1982), 19-45. 
\bibitem{s} S.~Salamon, {\it Riemannian geometry and holonomy groups},
Pitsman Res.\ Notes in Math.\ 201, Longman, Harlow 1989.
\bibitem{gh}A.~Gray and L.~Hervella, ``The sixteen classes of almost Hermitian
manifolds and their linear invariant'',
Ann.\ di Mat.\ Pura ed Appl.\ (IV), {\bf 123} (1980), 35-58. 
\bibitem{cs}S.~Chiossi and S.~Salamon, ``The intrinsic torsion of SU(3) and
G$_2$ structures,'' Proc.\ conf.\ Differential Geometry Valencia 2001, 
math.DG/0202282.

\end{thebibliography}
\end{document}